\documentclass[singlecolumn,12pt]{asme2ej}

\usepackage{epsfig} 
\usepackage{graphicx}
\usepackage[caption=false]{subfig}
\usepackage{amsmath}
\usepackage{float}
\usepackage{placeins}

\title{Pressure strain correlation modeling for turbulent flows}

\author{J. P. Panda\thanks{Address all correspondence to this author.}
    \affiliation{
	Department of Ocean Engineering and Naval Architecture,\\
IIT Kharagpur,\\
Kharagpur 721302, India\\
e-mail:  jppanda@iitkgp.ac.in
    }	
}

\author{H. V. Warrior
 \affiliation{
	Department of Ocean Engineering and Naval Architecture,\\
IIT Kharagpur,\\
Kharagpur 721302, India\\
e-mail:  warrior@naval.iitkgp.ernet.in
    }	

}

\begin{document}

\maketitle    

\begin{abstract}
{\it 
Almost all investigations of turbulent flows in academia and in the industry utilize some degree of turbulence modeling. Of the available approaches to turbulence modeling Reynolds Stress Models have the highest potential to replicate complex flow phenomena. Due to its complexity and its importance in flow evolution modeling of the pressure strain correlation mechanism is generally regarded as the key challenge for Reynolds Stress Models. In the present work, the modeling of the pressure strain correlation for complex turbulent flows is reviewed. Starting from the governing equations we outline the theory behind models for both the slow and rapid pressure strain correlation. Established models for both these are introduced and their successes and shortcomings are illustrated using simulations and comparisons to experimental and numerical studies. Recent advances and developments in this context are presented. Finally, challenges and hurdles for pressure strain correlation modeling are outlined and explained in detail to guide future investigations.   
}
\end{abstract}

\begin{nomenclature}

\entry{$R_{ij}$}{Reynolds stress tensor}
\entry{$b_{ij}$}{Reynolds stress anisotropy}
\entry{$P_{ij}$}{Production tensor}
\entry{$P$}{Turbulent kinetic energy production}
\entry{$\eta_{ij}$}{Dissipation tensor}
\entry{$\eta$}{Turbulent kinetic energy dissipation}
\entry{$T_{ijk,k}$}{Diffusive transport term}
\entry{$\phi_{ij}$}{Pressure strain correlation}
\entry{$\phi_{ij}^S$}{Slow pressure strain correlation}
\entry{$\phi_{ij}^R$}{Rapid pressure strain correlation}
\entry{$k$}{turbulent kinetic energy}
\entry{$q=2\sqrt{k}$}{turbulent velocity scale}
\entry{$\delta_{ij}$}{Kronecker delta tensor}
\entry{$\rho$}{density}
\entry{$\nu$}{Kinematic viscosity}
\entry{$U_i$}{Mean velocity component}
\entry{$u_i$}{Fluctuating velocity component}
\entry{$P$}{Mean pressure}
\entry{$p$}{Fluctuating pressure}
\end{nomenclature}

\section{Introduction}
Turbulent flows appear in problems of interest to many fields of engineering sciences such as aeronautics, mechanical, chemical engineering and in oceanographic, meteorological and astrophysical sciences, besides others. Improved understanding of turbulence evolution would lead to important advances in these fields. 

In academic and industrial applications, most investigations into turbulent flow problems use turbulence models. Turbulence models are simplified relations that express quantities that are difficult to compute in terms of simpler flow parameters. They relate higher-order unknown correlations to lower-order quantities. These unknown correlations represent the actions of viscous dissipation, pressure-velocity interactions, etc. For example pressure strain correlation is a non-local phenomenon and is difficult to compute. Using models for pressure strain correlation, it is expressed as a function of Reynolds stresses, dissipation and mean velocity gradients which are local quantities. This enables us to estimate the pressure strain correlation and its effects on flow evolution in a simpler manner that is computationally inexpensive. Turbulence models are an essential component of all computational fluid dynamics software and are used in almost all simulations into real life fluid flows of engineering importance. 

A majority of industrial applications use simple two-equation turbulence models like the $k-\epsilon$ and $k-\omega$  models. However recent emphasis in the scientific community has markedly shifted to Reynolds stress models (\cite{hanjalic2011,durbin2017,klifi2013,mishra3,jakirlic2015,manceau2015,eisfeld2016,schwarzkopf2016,moosaie2016,lee2016,mishra6,sun2017}). Reynolds stress models have the potential to provide better predictions than turbulent viscosity based models at a computational expense significantly lower than DNS studies. They may be able to model the directional effects of the Reynolds stresses and additional complex interactions in turbulent flows (\cite{johansson1994}). They have the ability to accurately model the return to isotropy of decaying turbulence and the behavior of turbulence in the rapid distortion limit (\cite{pope2000}).

Reynolds Stress Models are based on the Reynolds Stress Transport Equation that describes the evolution of individual components of the Reynolds stress tensor. This is in contrast to two-equation modeling approach where evolution equations for scalars like the turbulent kinetic energy $k$ and dissipation $\epsilon$ are solved and the eddy viscosity hypothesis is used to approximate the Reynolds stresses. The Reynolds Stress Transport Equations describe the production, dissipation and redistribution each of the components of the Reynolds stress tensor. Different physical mechanisms in this evolution are represented by the separate terms in this equation. The general form of the Reynolds Stress Transport Equation is given by \cite{pope2000}
\begin{equation}
\begin{split}
&\partial_{t} \overline{u_iu_j}+U_k \frac{\partial \overline{u_iu_j}}{\partial x_k}=P_{ij}-\frac{\partial T_{ijk}}{\partial x_k}-\eta_{ij}+\phi_{ij},\\
&\mbox{where},\\ 
& P_{ij}=-\overline{u_ku_j}\frac{\partial U_i}{\partial x_k}-\overline{u_iu_k}\frac{\partial U_j}{\partial x_k},\\
&\ T_{kij}=\overline{u_iu_ju_k}-\nu \frac{\partial \overline{u_iu_j}}{\partial{x_k}}+\delta_{jk}\overline{ u_i \frac{p}{\rho}}+\delta_{ik}\overline{ u_j \frac{p}{\rho}},\\
&\eta_{ij}=-2\nu\overline{\frac{\partial u_i}{\partial x_k}\frac{\partial u_j}{\partial x_k}}  \\
&\phi_{ij}= \overline{\frac{p}{\rho}(\frac{\partial u_i}{\partial x_j}+\frac{\partial u_j}{\partial x_i})}\\
\end{split}
\end{equation}
The turbulence production process is represented by $P_{ij}$ and is an inner product between the Reynolds stress tensor and the mean velocity gradient. In physical terms this mechanism models the action of the mean velocity gradients working against the Reynolds stresses and is a transfer of kinetic energy from the mean flow to the fluctuating velocity field. The production mechanism acts as a source of energy for the turbulent flow. $\eta_{ij}$ represents the dissipation process and is the product of the fluctuating velocity gradients and the fluctuating rate of strain. Physically it models the fluctuating velocity gradients working against the deviatoric fluctuating stresses transforming turbulence kinetic energy into internal energy. The dissipation mechanism acts as a sink of energy for the turbulent flow. The turbulent transport process is represented by $T_{ijk}$ and represents the transfer of turbulent kinetic energy between different locations in the flow domain. This has contributions from viscous diffusion, pressure transport and turbulent convection. Finally $\phi_{ij}$ represents the pressure strain correlation and redistributes turbulent kinetic energy among the components of the Reynolds stresses. Of these terms, production is the only process that is closed at the single point level. The other terms require models for their closure. The accuracy of the Reynolds stress modeling approach depends on the quality of the models developed for these turbulence processes.

Of the terms that require models for their closure the pressure strain correlation term is generally considered to be the most important \cite{lumley1975,sc}. There are three reasons behind this. Firstly the pressure strain correlation term is active in all turbulent flows. For instance in homogeneous turbulence, turbulence transport is absent due to spatial homogeneity. Similarly in decaying turbulence, turbulence production is zero due to the absence of mean velocity gradients. In the rapidly distorted turbulent flows, the dissipation mechanism is negligible as its time scale is much larger than the applied distortion. However in all these flows, the pressure strain correlation is present and actively transforming the evolution of the turbulent flow.

The second reason is due to the action of pressure strain correlation being very important in the evolution of turbulent flows. In important flow regimes like elliptic streamline flow the flow instability is initiated by pressure action \cite{kerswell2002}. In strained mean flows like plane strain, axisymmetric strained mean flows pressure action stabilizes the flow instability \cite{mishra4}. The pressure strain correlation term determines if turbulence grows or decays in many turbulent flows. Hence its accurate modeling is highly important.

The final reason is due to the complexity of the pressure strain correlation mechanism and the challenges in its modeling. The central ideas for pressure strain correlation modeling were introduced by \cite{chou}. The first model for the pressure strain correlation was formulated by \cite{rotta1951}. Since their foundational investigations, many researchers have developed more advanced and complex closure models for the pressure strain correlation term \cite{lrr, ssg,johansson1994,kr1995}. However all the available pressure strain correlation models have notable shortcomings. These shortcomings exist in the accuracy of their predictions and the realizability of their predictions. For example in rotation dominated flows the predictions of available models is incorrect both quantitatively and qualitatively. While Direct Numerical Simulations show that turbulence should be growing, available models predict that turbulence decays for these flows. Such rotation dominated flows include many flows of aerospace engineering like the trailing vortex, flap edge vortices and leading edge vortex flows. Another type of flows where the available pressure strain correlation
models are unsatisfactory is non-equilibrium turbulent flows. Non equilibrium turbulent flows include highly strained flows for example flows with shock turbulence interactions. The shortcomings of available models in accurately predicting such important classes of engineering flows is limits engineering investigations in such flows. In addition to accuracy in predictions the available pressure strain correlation models are unable to provide realizable predictions. The realizability condition \cite{realizability2, realizability1} tries to guarantee that the predictions of the turbulence model are not unphysical and correspond to flows that can exist in real life. In mathematical terms, the realizability condition requires turbulence models to predict a positive semi-definite Reynolds stress tensor. Unrealizable turbulence models may lead to issues in numerical convergence and numerical instability. Many investigators have found that the available pressure strain correlation models can guarantee realizable predictions only for low to moderate levels of Reynolds stress anisotropy \cite{mishra3}.  

The rest of this paper is organized as follows. In Section II we outline the mathematical details of pressure strain correlation modeling. In Section III we discuss the action of the slow pressure strain correlation term, introduce established models for this term and compare their performance for different flows. In Section IV we discuss the action of the rapid pressure strain correlation term, introduce established rapid pressure strain correlation models and compare their performance for different flows. In the first four sections we focus on outlining developments in the modeling of the pressure strain correlation for incompressible flows. In Section V we outline and discuss important challenges and hurdles for making pressure strain correlation models an indispensable tool in the engineering design process. This paper concludes with a summary and concluding remarks in Section VI.

\section{Mathematical and modeling details}

In incompressible flows fluctuating pressure is governed by a Poisson equation:
\begin{equation}
\frac{1}{\rho}{\nabla}^2 p=-2\frac{\partial{U}_j}{\partial{x}_i}\frac{\partial{u}_i}{\partial{x}_j}-\frac{\partial^2}{\partial x_i \partial x_j} ( u_iu_j-\overline{u_iu_j})
\end{equation}

Poisson equation is an elliptic partial differential equation. The Laplacian $\nabla^2$ is an elliptic operator and because of this the Poisson equation has no real characteristic directions. This elliptic nature of the operator leads to the non-local nature of the solution for fluctuating pressure. This elliptic nature of the governing equation indicates that the pressure at a single point in the flow is affected by changes to the flow at all points in the flow domain. This non-local character is inherited by the pressure strain correlation as well.

In literature this fluctuating pressure is decomposed into two components, rapid and slow pressure.
\begin{equation}
p=p^S+p^R
\end{equation}

Rapid pressure corresponds to the linear part of the source term in the Poisson equation. This term is directly and instantaneously affected by any changes in the mean velocity gradient and is referred to as rapid pressure. It is governed by
\begin{equation}
\frac{1}{\rho}{\nabla}^2 p^R=-2\frac{\partial{U}_j}{\partial{x}_i}\frac{\partial{u}_i}{\partial{x}_j}
\end{equation}

Slow pressure corresponds to the nonlinear part of the source term in the Poisson equation. This term is not directly affected by changes in the mean velocity gradient and is referred to as slow pressure. It is governed by
\begin{equation}
\frac{1}{\rho}{\nabla}^2 p^S=-\frac{\partial^2 }{\partial x_i \partial x_j} ( u_iu_j-\overline{u_iu_j})
\end{equation}
The most general form of the slow pressure strain correlation has the form \cite{ssmodel}:
\begin{equation}
\phi^{(S)}_{ij}=\beta_1 b_{ij} + \beta_2 (b_{ik}b_{kj}- \frac{1}{3}II_b \delta_{ij})
\end{equation}
$\beta_1$ and $\beta_2$ can be the functions of second and third invariants of Reynolds stress anisotropy or can be a function of turbulent Reynolds number. $b_{ij}=\frac{\overline{u_iu_j}}{2k}-\frac{\delta_{ij}}{3}$ is the Reynolds stress anisotropy tensor, $II_b$ and $III_b$ are the second and third invariants of the Reynolds stress anisotropy respectively.

The most general form of the rapid pressure strain correlation is:
\begin{equation}
\begin{split}
&\phi_{ij}^{(R)}=S_{pq}[Q_1\delta_{ip}\delta_{jq}+Q_2(b_{ip}\delta{jq}+b_{jp}\delta{iq-2/3b_{pq}}\delta{ij})+Q_3b_{pq}b_{ij}+Q_4(b_{iq}b_{jp}-1/3b_{pk}b_{kq}\delta_{ij})\\
&+Q_5b_{pl}b_{lq}b_{ij}+(Q_5b_{pq}+Q_6b_{pl}b_{lq})(b_{ik}b_{kj}-1/3II_{b}\delta_{ij}]\\
&+\Omega_{pq}[Q_7(b_{ip}\delta{jq}+b_{jp}\delta_{iq})+Q_8b_{pk}(b_{jk}\delta{iq}+b_{ik}\delta{jq}+Q_9b_{pk}(b_{jk}b_{ik}+b_{ik}b_{jq})]
\end{split}
\end{equation}
where, $S_{ij}$ is the mean rate of strain, $W_{ij}$ is the mean rate of rotation and $K$ is the turbulent kinetic energy. $II_b = b_{ij}b_{ji}$ is the second invariant of the Reynolds stress anisotropy tensor. Different models differ in the choice of the values of the model coefficients. Choosing specific sets of coefficients to be non-zero determines the order of the model with respect to the Reynolds stress tensor. This is outlined in Table 1.

\begin{table}\scriptsize
  \begin{center}
\def~{\hphantom{0}}
  \begin{tabular}{c c c}
\textbf{}     \textbf{RPSC Model}&\ \textbf{Expression Order}&\  \textbf{Non-zero Coefficients}\\[3pt]
      LRR \cite{lrr}   \ \    &\ \ \ \ linear in Reynolds stresses														  &\ \ \ \ $Q_1, Q_2, Q_7$\\ \\
   SSG \cite{ssg} \ \   &\ \ \ \ quadratic in Reynolds stresses                                 						    &\ \ \ \ $Q_1, Q_2, Q_3, Q_7$\\ \\
    Johansson \& Hallback \cite{johansson1994} \ \  &\ \ \ \ fourth order in Reynolds stresses            										&\ \ \ $Q_1-Q_9$ \\ \\

  \end{tabular}
  \\
  \caption{Rapid pressure strain correlation models compared with respect to their non-zero coefficients and order of expression}
  \label{tab:comp}
  \end{center}
\end{table}

As is seen in the general model expressions given in Equations (6) and (7) both the slow and the rapid components of the pressure strain correlation are modeled using tensors like the Reynolds stress anisotropy ($b_{ij}$), mean rate of strain ($S_{ij}$), mean rate of rotation ($\Omega_{ij}$), etc. These tensors in the modeling basis are local tensors but the pressure strain correlation has non-local dynamics. Because of this inconsistency most pressure strain correlation models have serious shortcomings in the accuracy of their predictions and in maintaining realizability of their predictions. 

In the next sections we discuss the slow and rapid pressure strain correlation models respectively. We find a similar trend in their development where the first few models are simpler and attempt to replicate some basic details of the pressure strain correlation. The succeeding models attempt to address shortcomings by incorporating more complex expressions, such as higher degrees of nonlinearity with respect to the Reynolds stress anisotropy terms. The final models attempt to add additional tensors to the modeling basis that can admit missing non-local information to the model.



\section{Slow pressure strain correlation models}
Numerous experimental and numerical investigations into decaying turbulent flows \cite{lumley1977,warhaft1978,le1985} have observed that along with a decay in the turbulent kinetic energy the anisotropy of the Reynolds stresses reduces towards an isotropic state. This is also observed in experimental investigations \cite{choi1984,choi2001,hallback1995} that initially anisotropic Reynolds stresses relax towards an isotropic state in the absence of external mean velocity gradients. This is known as the return to isotropy phenomenon of turbulence. In turbulence modeling the slow pressure strain correlation is chiefly responsible for the return to isotropy of turbulence. Slow pressure strain correlation models aim to replicate the details of this return to isotropy phenomenon. 


\subsection{Rotta Model:}
The first model for the slow pressure strain correlation was proposed by \cite{rotta1951}. The form of this model is given by
\begin{equation}
\phi^{(S)}_{ij}=-2C_{R}\epsilon b_{ij}
\end{equation}
The evolution equation for the Reynolds stress anisotropy for the Rotta model can be written as \cite{pope2000}
\begin{equation}
\frac{db_{ij}}{dt}=(-C_{R}-1)\frac {\epsilon} {k}  b_{ij}
\end{equation}

The slow pressure strain correlation model of Rotta \cite{rotta1951} is linear in the Reynolds stresses. While it captures the return to anisotropy it is unable to capture the nonlinear nature of this return to isotropy process. For example on the Lumley triangle the paths predicted by the Rotta model are straight lines. Experimental data clearly shows that the return to isotropy is via curved trajectories \cite{chung1995}. The dependence of the rate of return to isotropy on the invariants of the Reynolds stress anisotropies is also not accounted for in the Rotta model. 


\subsection{Lumley Model:}
The nonlinear effects were incorporated in the model of \cite{lumley1979}, in which the nonlinearities were introduced in the model through the functions of Reynolds stress anisotropy or the invariants of the Reynolds stress anisotropy. The coefficients of the model were taken as the function of turbulence Reynolds number 
\begin{equation}
\begin{split}
\beta_1=2.0+\frac{F}{9}\exp\frac{-7.7}{\sqrt{Re_t}}\Bigg(\frac{72}{\sqrt{Re_t}}+80.1  ln[1+62.4(-II_b+2.3III_b)]\Bigg),and\\
\beta_2=0
\end{split}
\end{equation}

where, $q^2=2K$, $II_b$ and $III_b$ are the second and third invariants of the Reynolds stress anisotropy tensor. $F$ is the determinant of the normalized Reynolds stress tensor. 

\subsection{Sarkar and Speziale Model:}
The model of Sarkar and Speziale \cite{ssmodel} is a quadratic model, the coefficients of the model are constants. 

\begin{equation}
\beta_1=3.4 \quad and \quad \beta_2=3(\beta_1-2)
\end{equation}
The transport equations for the Reynolds stress anisotropy were considered as follows:
\begin{equation}
\begin{split}
\frac{dII_b}{d\tau}=1.4II_b+8.4III_b\\
\frac{dIII_b}{d\tau}=-4.2III_b+2.1II_b^2
\end{split}
\end{equation}

This simple quadratic model with only one independent constant is able to account for most of the nonlinear character of the return to isotropy phenomenon. It has been widely adopted in engineering simulations of turbulent flows.

\subsection{Slow pressure strain correlation models with extended tensor bases:}
All the slow pressure strain correlation models discussed in this section assume that the characteristic length scale of turbulence is the same in all directions. This is markedly true in flows where the geometry of the flow domain or body forces lead to a co-ordinate direction in the flow being decidedly preferred. For example axisymmetric expansion and axisymmetric contraction mean flows. In many anisotropic turbulent flows, the characteristic length scale is observed to be varying in different directions \cite{panda2017}. At the most basic level, we must try to include this anisotropy in the length scale in the modeling basis for the pressure strain correlation. \cite{panda2017} proposed a modeled formulation of length scale anisotropy tensor by assuming a linear relationship of length scale anisotropy with dissipation and Reynolds stress anisotropy tensors. The length scale anisotropy \cite{warrior2014} has the form 
\begin{equation}
L_{ij}/l=0.75(c_1^*b_{ij}+c_2^*e_{ij})
\end{equation}
The model for the slow pressure strain correlation has the form:
\begin{equation}
\begin{split}
\phi_{ij}=c_1b_{ij}+c_2e_{ij}+c_3l_{ij}+c_4(b_{ik}b_{kj}-1/3b_{mn}b_{mn}\delta_{ij})\\
+c_5(b_{ik}e_{kj}-1/3b_{mn}e_{mn}\delta{ij})+c_6(e_{ik}e_{kj}-1/3e_{mn}e_{mn}\delta{ij})+\\
c_7(b_{ik}l_{kj}-1/3b_{mn}l_{mn}\delta{ij})+c_8(e_{ik}l_{kj}-1/3e_{mn}l_{mn}\delta{ij})+\\
c_9(l_{ik}l_{kj}-1/3l_{mn}l_{mn}\delta{ij})
\end{split}
\end{equation}
The anisotropy states in the turbulent flows can be characterized by two variables $\xi$ and $\eta$ given by 
\begin{equation}
\begin{split}
\xi=(III_b/2)^{1/3}, \quad \eta=(-II_b/3)^{1/2},
\end{split}
\end{equation}
In a turbulent flow , $\xi$ and $\eta$ can be determined at any point and time from the Reynolds stresses. The $\xi$-$\eta$ phase space is bounded by two straight line segments denoting axisymmetric turbulence and from above by a curved line representing two-dimensional turbulence. This representation is referred to as the Lumley triangle \cite{pope2000} and is a simple manner to visualize the anisotropy of the Reynolds stress tensor. All realizable states of the Reynolds stress tensor lie inside the Lumley triangle. 

We compare the predictions of the slow pressure strain correlation models introduced in this section and contrast them to experimental data. Here we focus on three models specifically: the slow pressure strain correlation models of \cite{rotta1951}, \cite{ssmodel} and \cite{panda2017}. The model of \cite{rotta1951} is chosen as due to its simplicity it is widely used in turbulence simulations. The model of \cite{ssmodel} attempts to address the deficiencies of the linear \cite{rotta1951} model by adding nonlinear terms in the Reynolds stress anisotropy. This is one methodology to address the limitations in the models and is thus included. The model of \cite{panda2017} attempts to address the deficiencies of slow pressure strain correlation models by adding additional tensors to the modeling basis. This represents another approach to address the limitations in the models and is thus included in the comparisons. 

\begin{figure}
\centering
\includegraphics[height=8cm]{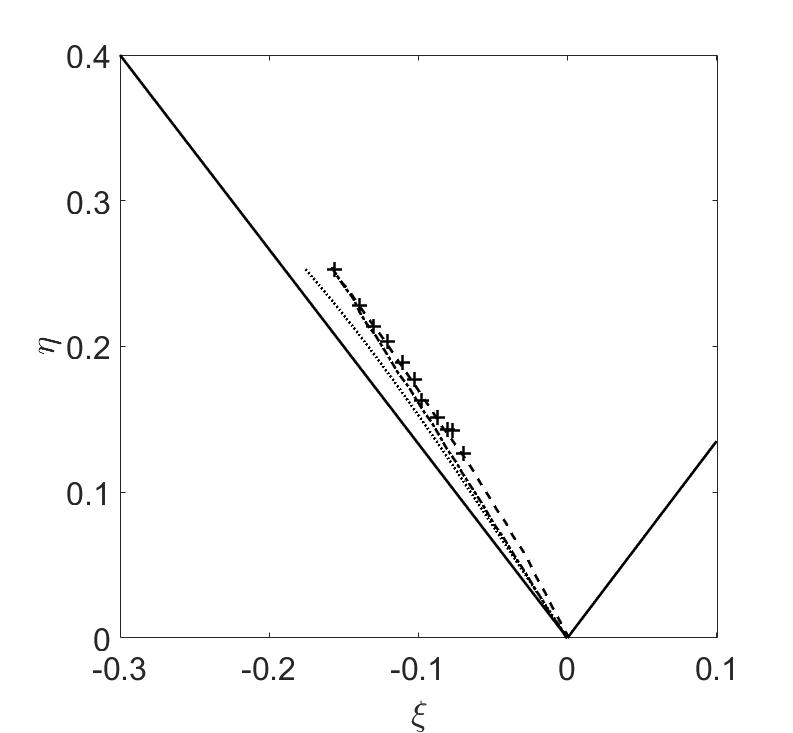}
\caption{Lumley triangle trajectories of the model predictions against the plane contraction experiment of \cite{le1985} ($III_b<0$) }
\end{figure}

\begin{figure}
\centering
\includegraphics[height=8cm]{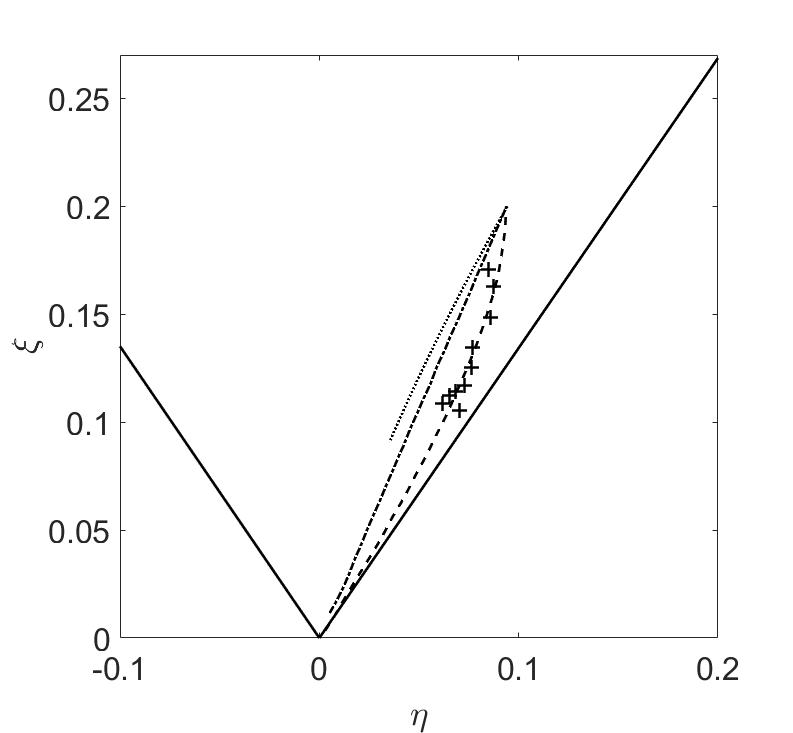}
\caption{Lumley triangle trajectories of  model predictions against the plane distortion experiment of \cite{choi1984}}
\end{figure}

\begin{figure}
\centering
\includegraphics[height=12cm]{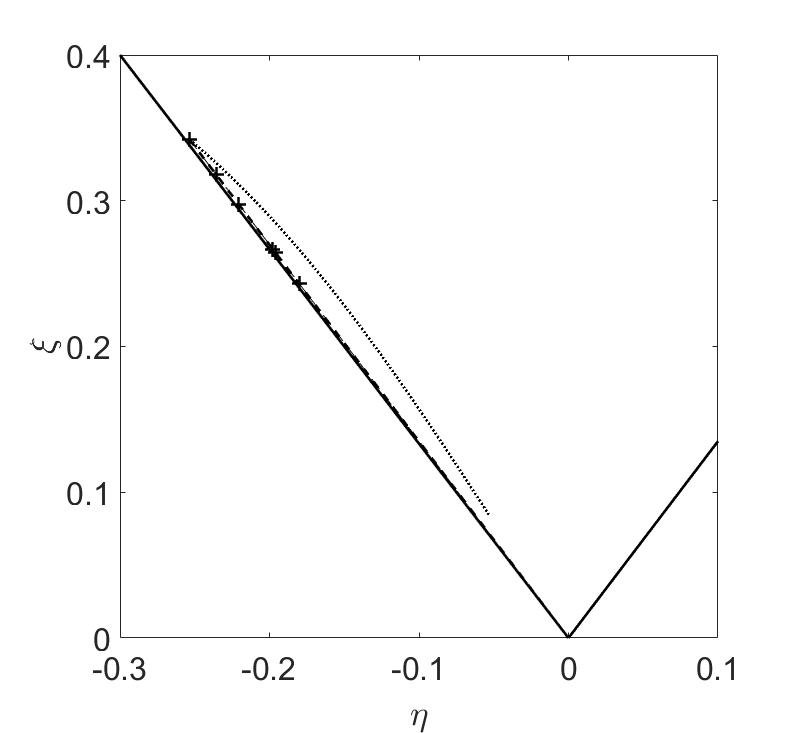}
\caption{Lumley triangle trajectories of model predictions against the experiment of \cite{warhaft1978}}
\end{figure}

Figure 1 represents the evolution of trajectories for the plane contraction experiment of \cite{le1985}, for this case the initial value of third invariant of Reynolds stress anisotropy is negative. The experimental data shows mild curvature in the phase plane. The nonlinear quadratic model of \cite{ssmodel} has predictions that better fit the experimental results, in comparison to \cite{rotta1951} and \cite{panda2017} models. The trajectories on the Lumley triangle in figure 2 are strongly curved indicating the nonlinear effects in the return to isotropy behavior. It is noticed in figure 2 that model of \cite{ssmodel} has predictions that are better in comparison other two models. The phase space comparison for the experiments of \cite{warhaft1978} is presented in figure 3. The curvature in the experimental results is very small. Both models of \cite{ssmodel} and \cite{rotta1951} have predictions that are very similar same. This indicates that the nonlinear effects were not dominant in the flows. 

\begin{figure}
\centering
\subfloat[$II_b$]{\includegraphics[height=8cm]{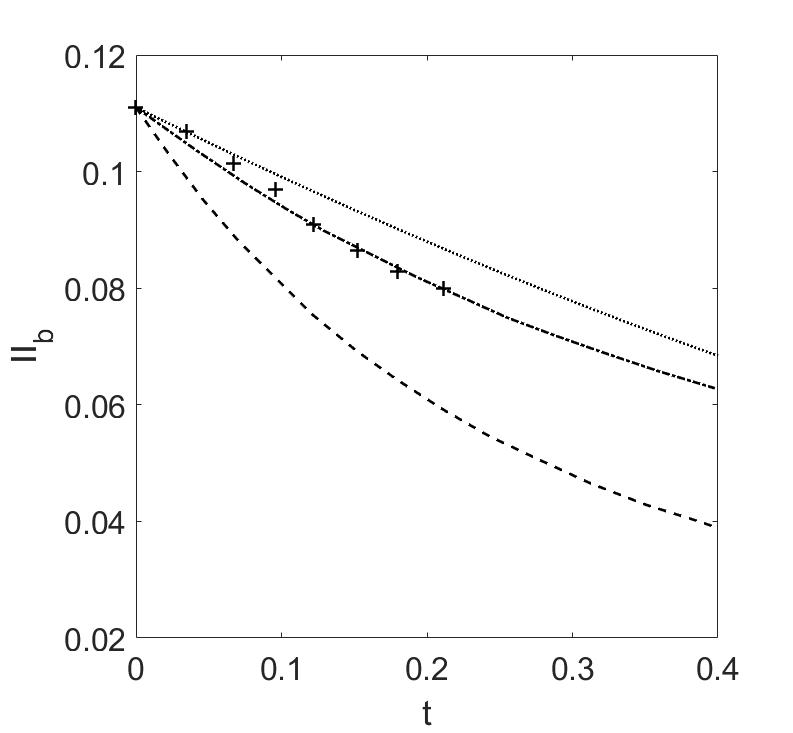}}
\subfloat[$III_b$]{\includegraphics[height=8cm]{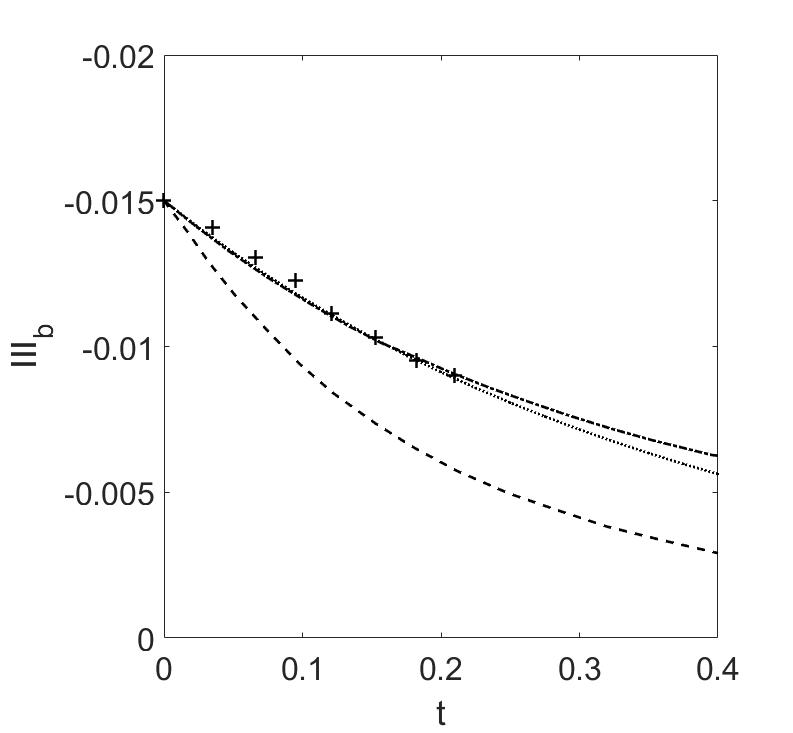}}
\caption{Comparisons of model predictions against the experiment of \cite{uberoi1963}}
\end{figure}

Figure 4 represents the temporal evolution of second ($II_b$) and third ($III_b$) invariants for the experimental results of \cite{uberoi1963}. The  \cite{panda2017} model has predictions that are better in comparison to the model of \cite{ssmodel}. 

From the comparison against experimental data the predictions of the model of \cite{ssmodel} show best agreement with data across different experimental studies.  

\section{Rapid pressure strain correlation models}

The rapid component of the pressure strain correlation accounts for the linear interactions between the fluctuating velocity field and the mean velocity gradient. This rapid pressure strain correlation has behavior that is very dependent on the mean velocity field. As an illustrative example we can demarcate the behavior of the rapid pressure strain correlation in two regimes of planar flows: hyperbolic streamline flow and elliptic streamline flow. In elliptic streamline flows the elliptical flow instability is initiated by the rapid pressure strain correlation \cite{kerswell2002}. In hyperbolic streamline flows the rapid pressure strain correlation stabilizes the flow instability \cite{mishra4}. The effect of the rapid pressure strain correlation is highly dependent on the mean gradient and substantially varies between different flows. 

Besides giving accurate predictions of the flow evolution and ensuring realizable Reynolds stresses there are additional properties required of the ideal RPSC model. These include
\begin{enumerate}
\item The RPSC model ($\phi_{ij}^{(R)}$) should have a model expression linear in the Reynolds stresses \cite{reynolds1976,pope2000}.
\item The RPSC model should have a model expression linear in the mean velocity gradient \cite{johansson1994, pope2000}.
\item The RPSC model should obey the Crow constraint (from isotropic initial conditions) \cite{crow1968}.
\end{enumerate}

An ideal model is expected to conform to these properties. However no available model is able to meet all these properties and still produce accurate predictions. While there are many available models for the rapid pressure strain correlation, we discuss three established popular models. These include the models by Launder, Reece and Rodi\cite{lrr} (termed the LRR model), Speziale, Sarkar and Gatski \cite{ssg} (termed the SSG model) and Johansson and Hallback \cite{johansson1994} (termed the Johansson-Hallback model).




\subsection{LRR Model:}
The model proposed by Launder, Reece and Rodi\cite{lrr} has the form
\begin{equation}
\begin{split}
\phi_{ij}^{(R)}=C_1K S_{ij}+C_2K(b_{ik} S_{jk}+b_{jk} S_{ik}-2/3b_{mn} S_{mn}\delta_{ij})+C_3K(b_{ik} W_{jk}+b_{jk} W_{ik})
\end{split}
\end{equation}
The closure coefficients are given as $C_1=0.8$, $C_2=1.75$ and $C_3=1.31$.
 
The model proposed by \cite{lrr} conforms to all the properties for a RPSC model. It is linear in the Reynolds stresses and the mean velocity gradient. It also conforms to the Crow constraint. However it is not able to show accurate predictions for complex flows for example flows dominated by rotational effects. It is also not able to maintain realizability of the Reynolds stress or their evolution for moderate to high levels of anisotropy in the flow \cite{mishra3}. 

The LRR model has been widely adopted in turbulence simulations and is available in most computational fluid dynamics software. Many variants of this model have also been developed. For example the model of \cite{jones1988} retains the model form of the LRR model but changes the coefficient values to improve performance in turbulent flows with high rates of shear.   

\subsection{SSG Model:}
The model proposed by Speziale, Sarkar and Gatski \cite{ssg} has the form
\begin{equation}
\begin{split}
\phi_{ij}^{(R)}=(C_1-C_1^*II^{0.5})K S_{ij}+C_2K(b_{ik} S_{jk}+b_{jk} S_{ik}-2/3b_{mn} S_{mn}\delta_{ij})+C_3K(b_{ik} W_{jk}+b_{jk} W_{ik})
\end{split}
\end{equation}
The closure coefficients are given as  $C_{1}=0.8$, $C_1^{*}=1.3$, $C_2=1.25$ and $C_3=0.4$.

The model expression does not conform to all the properties for RPSC models stated above and is quadratic in the Reynolds stresses. However it is able to show much improved accuracy in predictions and better realizability behavior than other linear models. For turbulent flows in non-inertial frames of reference this model is much better than other RPSC models.

\subsection {Johansson-Hallback Model:}
Johansson and Hallback\cite{johansson1994} derived the most general expression for the RPSC model. This is given by
\begin{equation}
\begin{split}
&\phi_{ij}^{(R)}=S_{pq}[Q_1\delta_{ip}\delta_{jq}+Q_2(b_{ip}\delta{jq}+b_{jp}\delta{iq-2/3b_{pq}}\delta{ij})+Q_3b_{pq}b_{ij}+Q_4(b_{iq}b_{jp}\\
&-1/3b_{pk}b_{kq}\delta_{ij})+Q_5b_{pl}b_{lq}b_{ij}+(Q_5b_{pq}+Q_6b_{pl}b_{lq})(b_{ik}b_{kj}-1/3II_{b}\delta_{ij}]\\
&+\Omega_{pq}[Q_7(b_{ip}\delta{jq}+b_{jp}\delta_{iq})+Q_8b_{pk}(b_{jk}\delta{iq}+b_{ik}\delta{jq}+Q_9b_{pk}(b_{jk}b_{ik}+b_{ik}b_{jq})]
\end{split}
\end{equation}

Here $Q_i$ are scalar functions of the invariants of the Reynolds stress anisotropy and the mean velocity gradient. These can in turn be expressed in terms of scalars $B_{\alpha}$ as 
\begin{equation}
\begin{split}
&Q_1=4/5-2/5(4B_2+15B_3)II_\alpha-2/5B_5III_\alpha-1/220(19B_6-120B_7)II_b^2,\\
&Q_2=-12B_1-1/2B_5II_b-1/2(B_6-8B_7)III_b,\\
&Q_3=-8B_2+36B_3+1/22(7B_6-72B_7)II_b,\\
&Q_4=96B_2-36B_3-1/22(7B_6-72B_7)II_b,\\
&Q_5=B_5,\\
&Q_6=B_6,\\
&Q_7=-4/3-28/3B_1+1/6(2B_4-B_5)II_b-1/18(3B_6-56B_7)III_b,\\
&Q_8=-16B_2+28B_3+1/22(3B_6-56B_7)II_b,\\
&Q_9=B_4
\end{split}
\end{equation}

Based on the choices for the $B_{\alpha}$ \cite{johansson1994,sjogren1998,sjogren2000,hallback1995} outlined models that were second-, third- and fourth-order with respect to the Reynolds stresses. All these models conform to the strong realizability condition. The fourth-order model shows high agreement with data from experiments for strain-dominated mean flows. The performance of this model for rotation-dominated mean flows is still lacking.

In figures 5, 6 and 7, we show the performance of these models in the rapid distortion limit by comparing them to the results of RDT simulations. For purely strained flows such as the plane strain mean flow shown in figure 5 the models have good agreement with the trends observed in the RDT simulation. The fourth-order model of \cite{johansson1994} shows better performance as compared to the models of \cite{lrr} and \cite{ssg}. 

This trend is observed again in figure 6 for a planar strained mean flow at the rapid distortion limit. All the models show acceptable agreement with the RDT simulation for the Reynolds stress anisotropy. The agreement of the fourth-order model of \cite{johansson1994} for the evolution of the turbulent kinetic energy is very accurate.

In figure 7 we compare the predictions of these models for an elliptic streamline flow at the rapid distortion limit. The figure shows that none of the models give satisfactory predictions for elliptic streamline flows. For example in figure 7 (b) the RDT simulations suggest that the turbulent kinetic energy of the flow is growing exponentially. All the RPSC models predict otherwise. For elliptic streamline flows RPSC models are inexact. 

\begin{figure}
\centering
\subfloat[$b_{11}$]{\includegraphics[height=8cm]{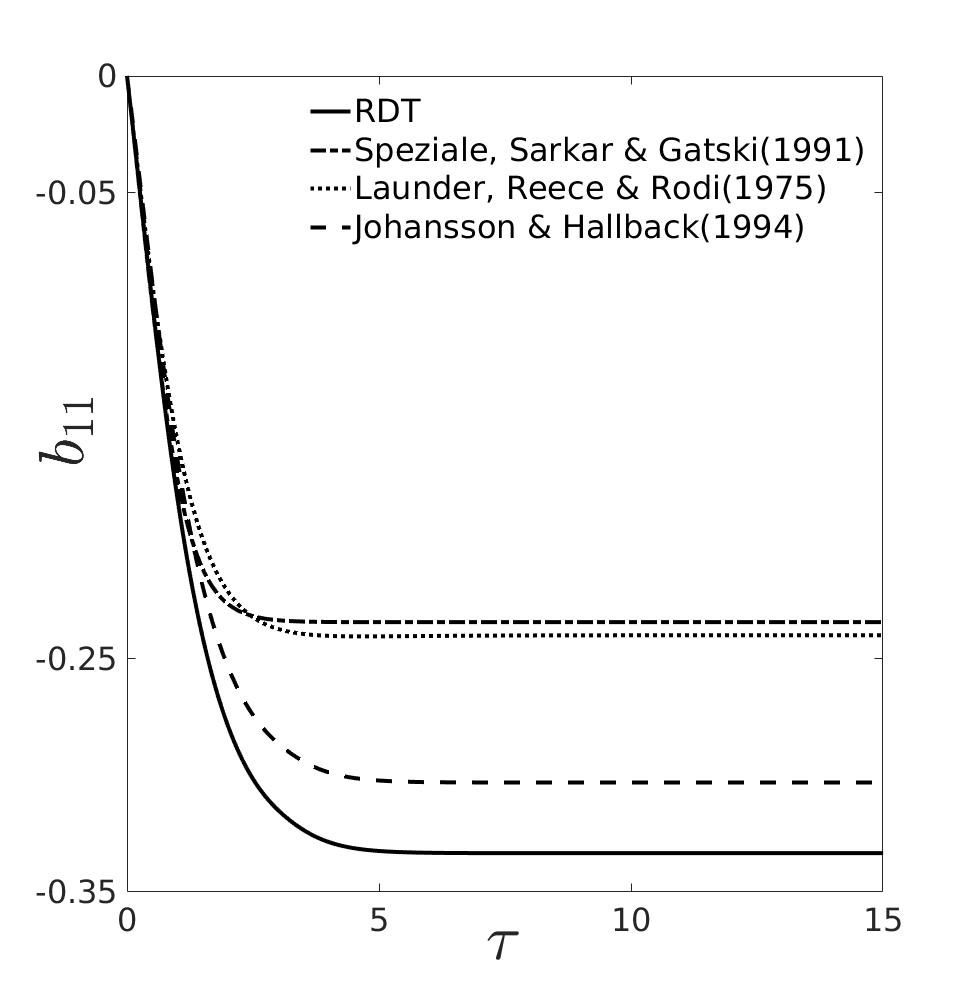}}
\subfloat[$log(k)$]{\includegraphics[height=8cm]{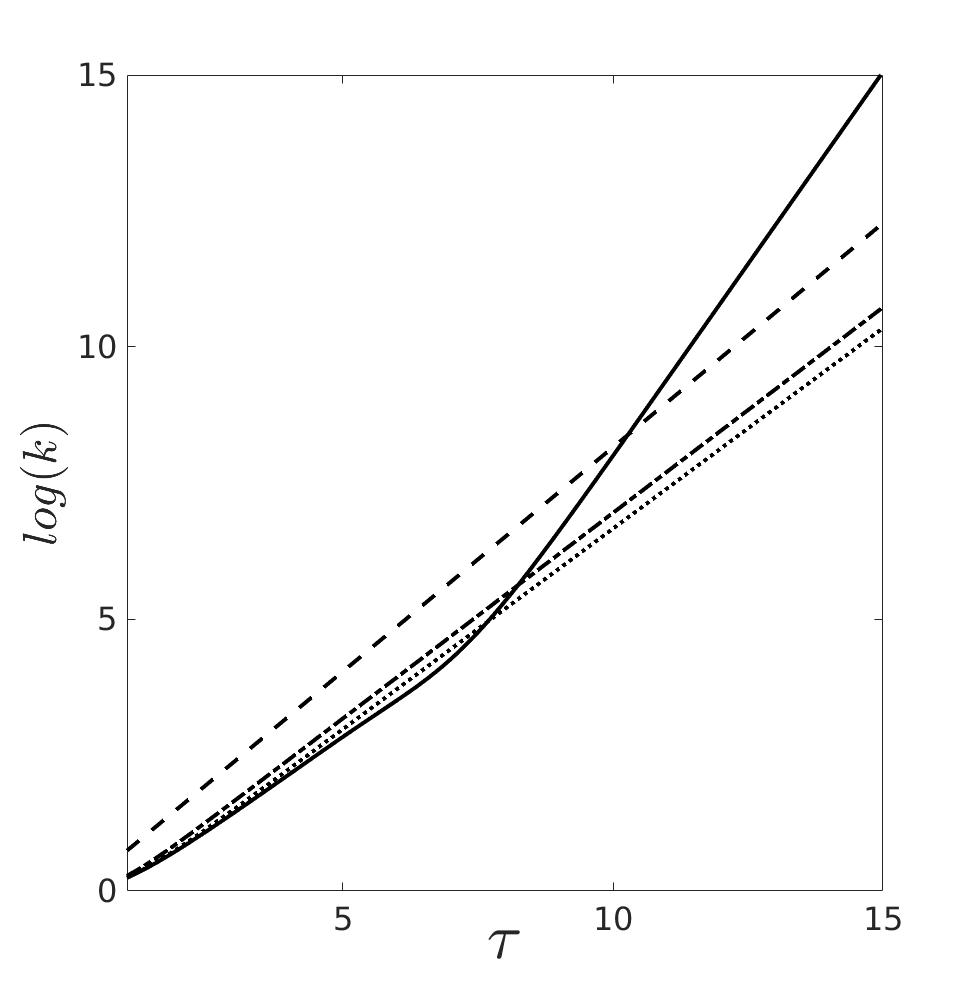}}
\caption{Rapid pressures train correlation model predictions for a plane strain mean flow at the rapid distortion limit}
\end{figure}

\begin{figure}
\centering
\subfloat[$b_{12}$]{\includegraphics[height=8cm]{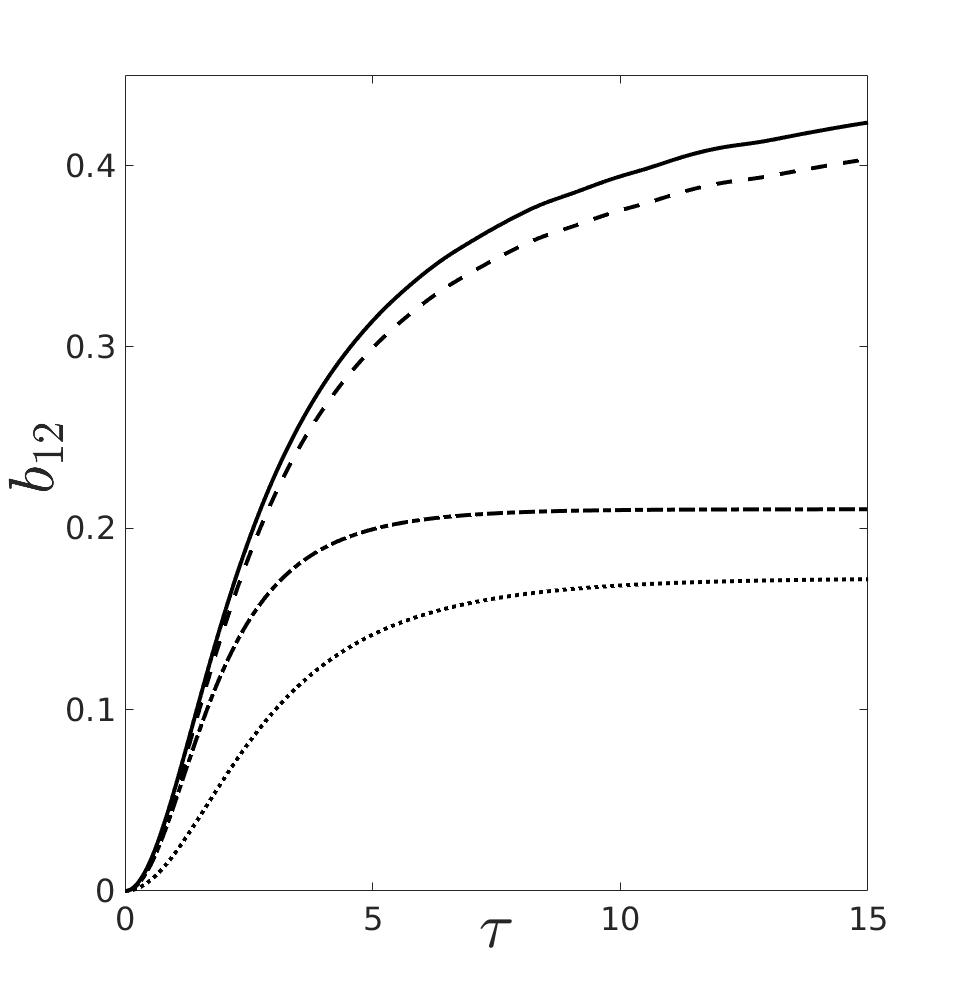}}
\subfloat[$log(k)$]{\includegraphics[height=8cm]{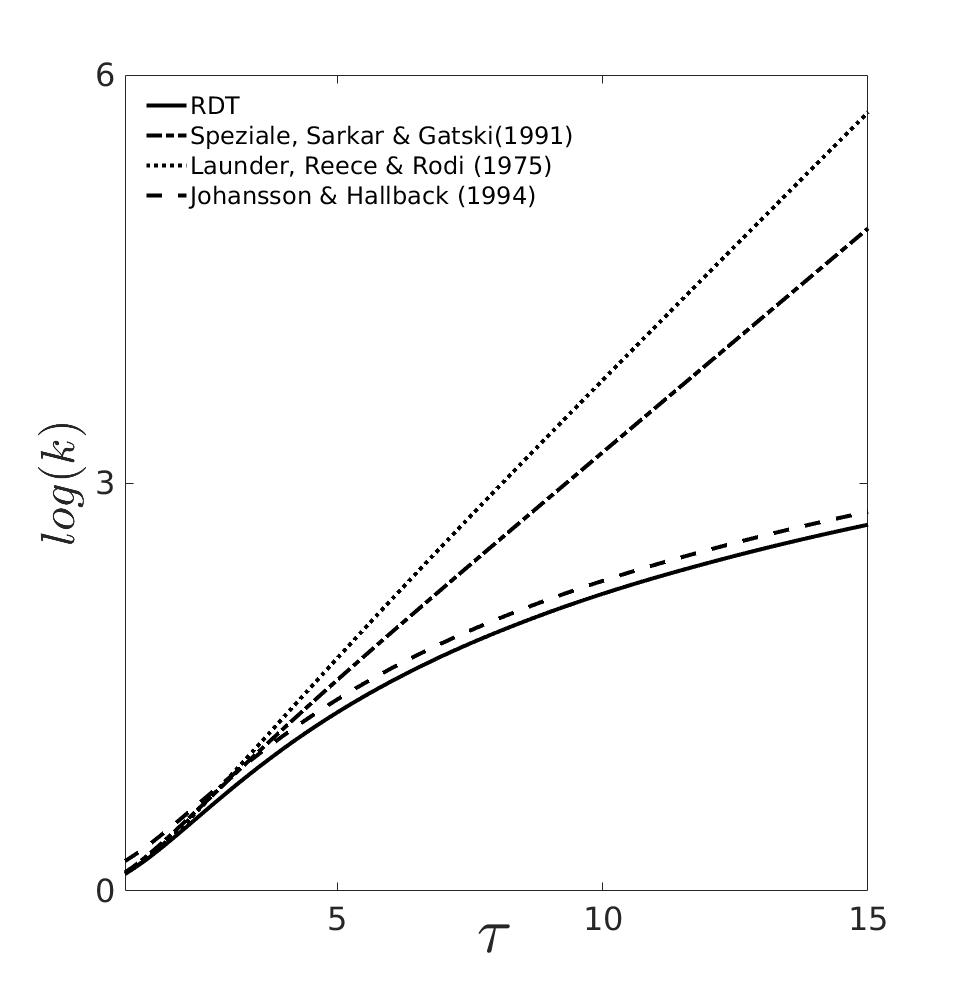}}
\caption{Rapid pressures train correlation model predictions for a planar strained mean flow at the rapid distortion limit}
\end{figure}

\begin{figure}
\centering
\subfloat[$b_{22}$]{\includegraphics[height=8cm]{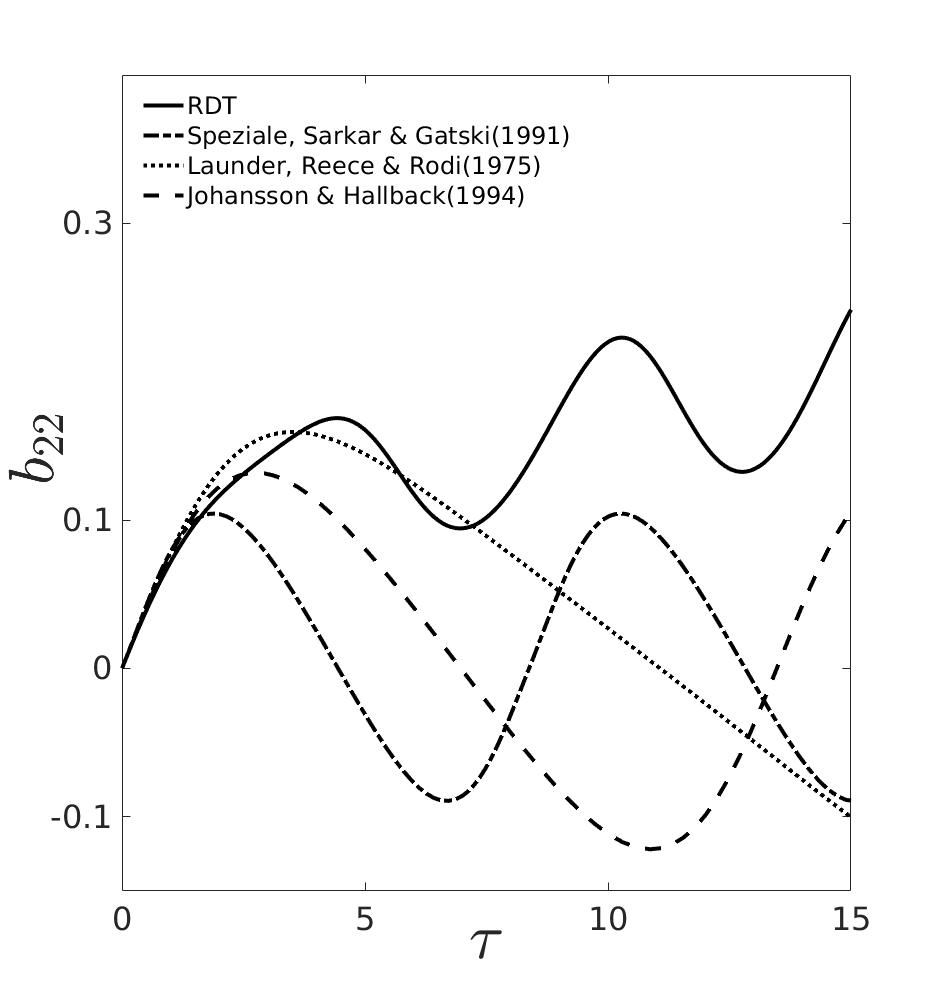}}
\subfloat[$log(k)$]{\includegraphics[height=8cm]{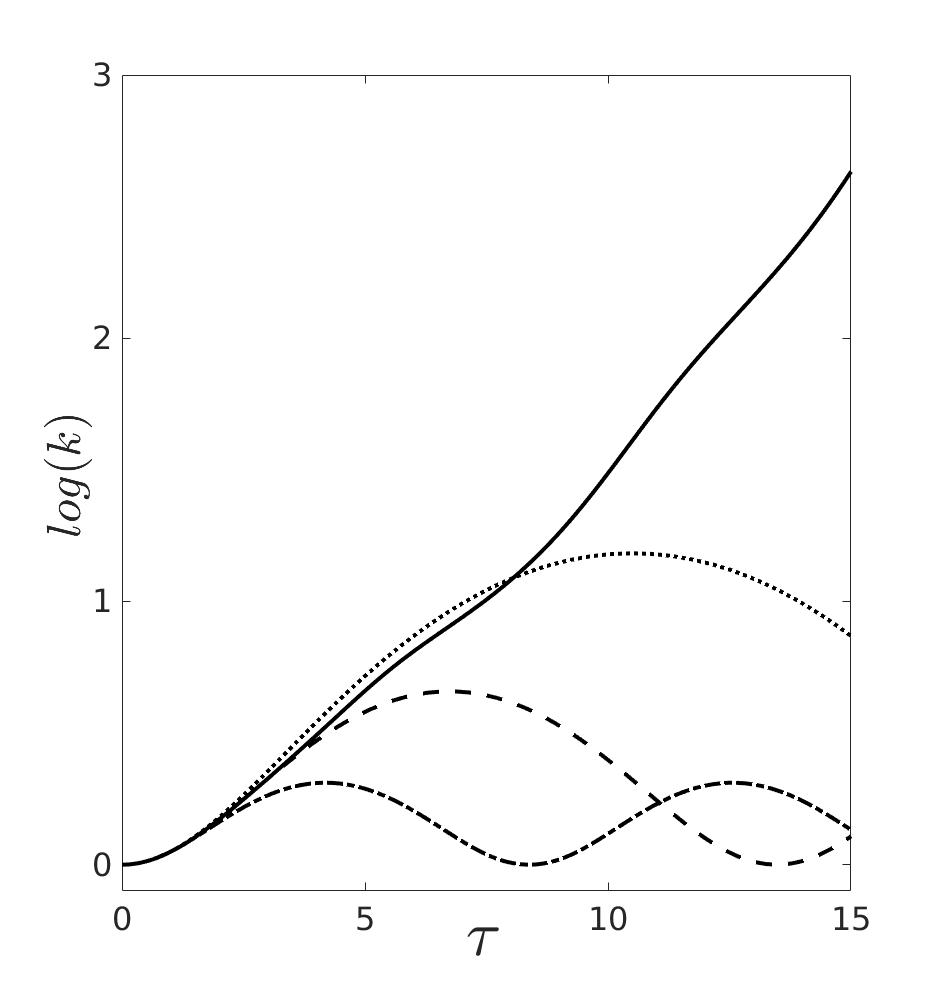}}
\caption{Rapid pressures train correlation model predictions for an elliptic streamline mean flow at the rapid distortion limit}
\end{figure}

\subsection {Rapid pressure strain correlation models with extended bases}
One of the primary challenges in RPSC modeling is to replicate the non-local dynamics of pressure while using local tensors such as the Reynolds stresses and mean velocity gradients. The models of \cite{lrr, ssg,johansson1994} attempt to do this but have unsatisfactory performance in rotation dominated flows, etc. Some investigators have tried to formulate RPSC models by appending additional tensors to the modeling basis. We discuss a few such notable models here and analyze one of these in detail.  

Kassinos and Reynolds\cite{kr1995} attempted to formulate a RPSC model using additional tensors in the modeling basis including as stropholysis, circulicity, etc. This was justified by differentiating between the componentiality and the
dimensionality of turbulent flow field. Using single point (or local) tensors such as the Reynolds stresses informs the model about the componentiality of the turbulent flow field but not about the dimensionality of the turbulent flow field. \cite{kr1995} define the structure dimensionality tensor $D_{ij}=M_{kkij}$, where the Reynolds stress is given by $R_{ij}=M_{ijkk}$. Addition of this tensor to the modeling basis would bring in important information and improve predictions. The final model in \cite{kr1995} did not show much improved predictions for rotation dominated flows. The model also had realizability issues \cite{mishra3}. The structure dimensionality tensor $D_{ij}$ is non-local and is not available in most engineering simulations. This made the usage of this model more problematic. 

Cambon et al\cite{cambon1992} posited that using just the deviatoric component of the Reynolds stresses was unable to describe the turbulent flow field completely especially in the presence of mean rotation. They decomposed the Reynolds stress anisotropy tensor into two components: directional and polarization anisotropy ($b_{ij}=b^e_{ij}+b^z_{ij}$). Transport equations for these two components separately were developed. This model was able to show some improvements in rotation dominated flows. The transport equations for the decomposed anisotropy components were not unique and the closure coefficients were tuned to give agreement with the experiments used in the investigation. Above all in a real life engineering problem there is no clear manner on how to decompose the Reynolds stress anisotropy as information about the decomposition is non-local.

\cite{mishra6} developed an illustrative model where the model closure coefficients were functions of the mean velocity gradient invariants. In previous investigations \cite{mishra1,mishra2} have illustrated the details of the intercomponent energy transfer caused by the rapid pressure strain correlation. Using  spectral analysis, \cite{mishra1,mishra2} establish a most likely evolution based on the statistics of the turbulent velocity that models should aim to reproduce. They have shown that including the mean velocity gradient in the modeling basis would lead to the addition of missing physics and improved model predictions. This is in agreement with \cite{lee1989} where it is shown that adding the mean strain rate information would improve the predictions of the pressure strain correlation model. In \cite{mishra3} a new approach to realizability is developed. Using this realizability approach it was shown that addition of the mean velocity gradient information would lead to better realizability behavior. \cite{mishra6,mishrathesis} have shown that including the mean gradient information by making the model coefficients functions of the mean velocity gradient would lead to a simple model structure, better realizability behavior and improved accuracy of predictions. 

Instead of adding non-local tensors to the modeling basis the model of \cite{mishra6} uses the mean velocity gradient invariants to add missing information to the model expression. This model may be considered as compliant to use in real life engineering problems as it does not require the estimation of non-local tensors. The model expression is given by
\begin{equation}
\begin{split}
\phi_{ij}^{(R)}=4/5K\overline S_{ij}+6A_5\beta S_{pq}K(b_{ip}\delta {jq} +b_{jp}\delta {iq} + 2/3b_{pq}\delta {ij})+\\
2/3(4+7A_5(\beta))W_{pq}(b_{ip}\delta{jq}+b_{jp}\delta{iq})
\end{split}
\end{equation}
\begin{equation}
\begin{split}
A_5(\beta)=0.22\beta -0.44,\beta\in [0,0.5] \quad and\\
A_5(\beta)=-0.83\beta^2-0.44,\beta\in [0.5,1]
\end{split}
\end{equation}
where ,$\beta$ is the ellipticity parameter and is defined as
\begin{equation}
\begin{split}
\beta=\frac{W_{mn}W_{mn}}{W_{mn}W_{mn}+s_{mn}s_{mn}}
\end{split}
\end{equation}

In figures 8 and 9 we compare the predictions of this model to the DNS investigation of \cite{bns} for different elliptic flows. The predictions of the LRR model are included for contrast. Blaisdell and Shariff \cite{bns} have simulated homogeneous turbulence subjected to elliptic mean flows:
\begin{equation}
\frac{\partial U_i}{\partial x_j}=
\begin{bmatrix}
  0 & 0 & -\gamma-e \\
  0 & 0 & 0 \\
  \gamma-e & 0 & 0
\end{bmatrix}
\end{equation}
where $e=\sqrt{\frac{1-\beta}{2}}$ and $\gamma=\sqrt{\frac{\beta}{2}}$. For $e>\gamma$ the mean flow has elliptic streamlines of aspect ratio $E=\sqrt{(\gamma+e)(\gamma-e)}$. We use this data from two simulations with mean flows having aspect ratios $E=2$ and $1.5$. The turbulent velocity field is initially isotropic and the initial $\frac{\eta}{Sk} = 0.167$. 

In figure 8 the case with $E=1.5$ is shown. In figure 8 (a) The model of \cite{mishra6} shows better agreement with DNS data than the predictions of the LRR model. In figure 8 (b) the LRR model does predict turbulent kinetic energy growth but at a rate much smaller than DNS. The rate of turbulent kinetic energy growth predicted by the model of \cite{mishra6} is in agreement with the DNS data. 

In figure 9 the case with $E=2$ is shown. (In this case the effect of rotation on flow evolution has increased over $E=1.5$). In figure 9 (b) while the DNS simulations predict the turbulent kinetic energy to be growing the model of LRR predicts decay. The rate of turbulent kinetic energy growth predicted by the model of \cite{mishra6} is in agreement with the DNS data.

\begin{figure}
\centering
\subfloat[$b_{22}$]{\includegraphics[height=8cm]{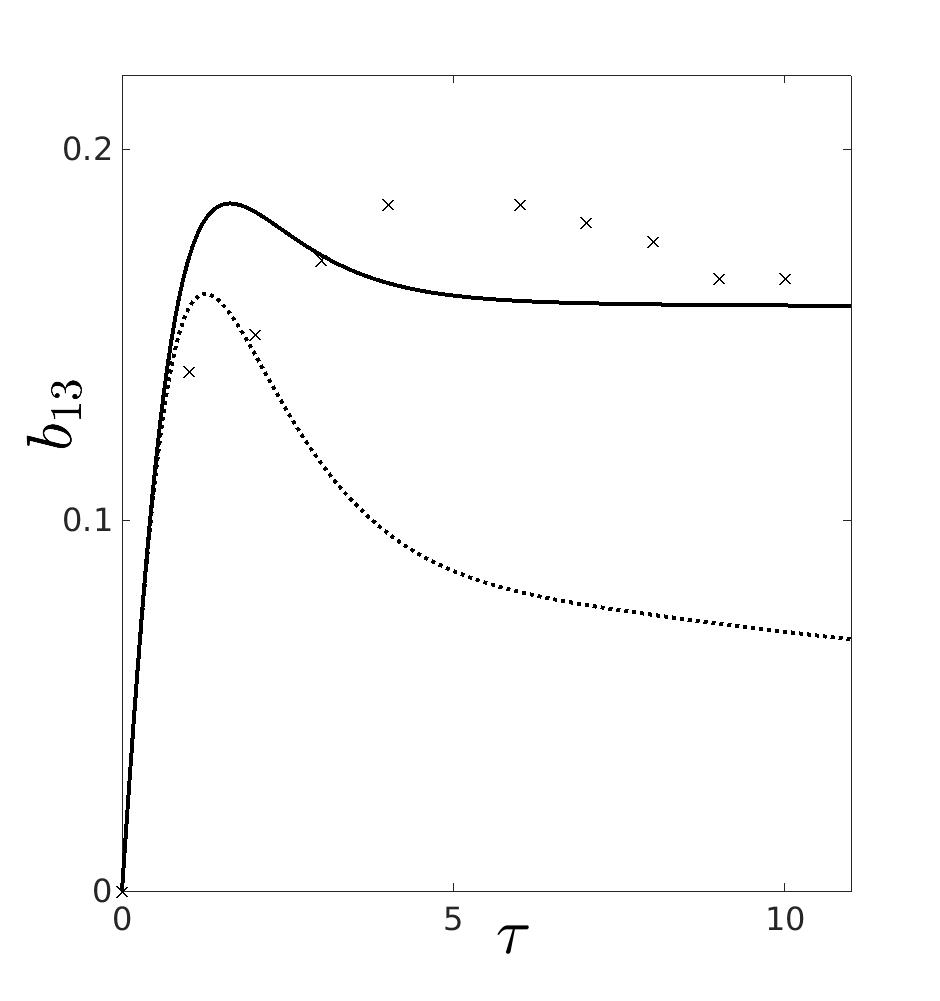}}
\subfloat[$log(k)$]{\includegraphics[height=8cm]{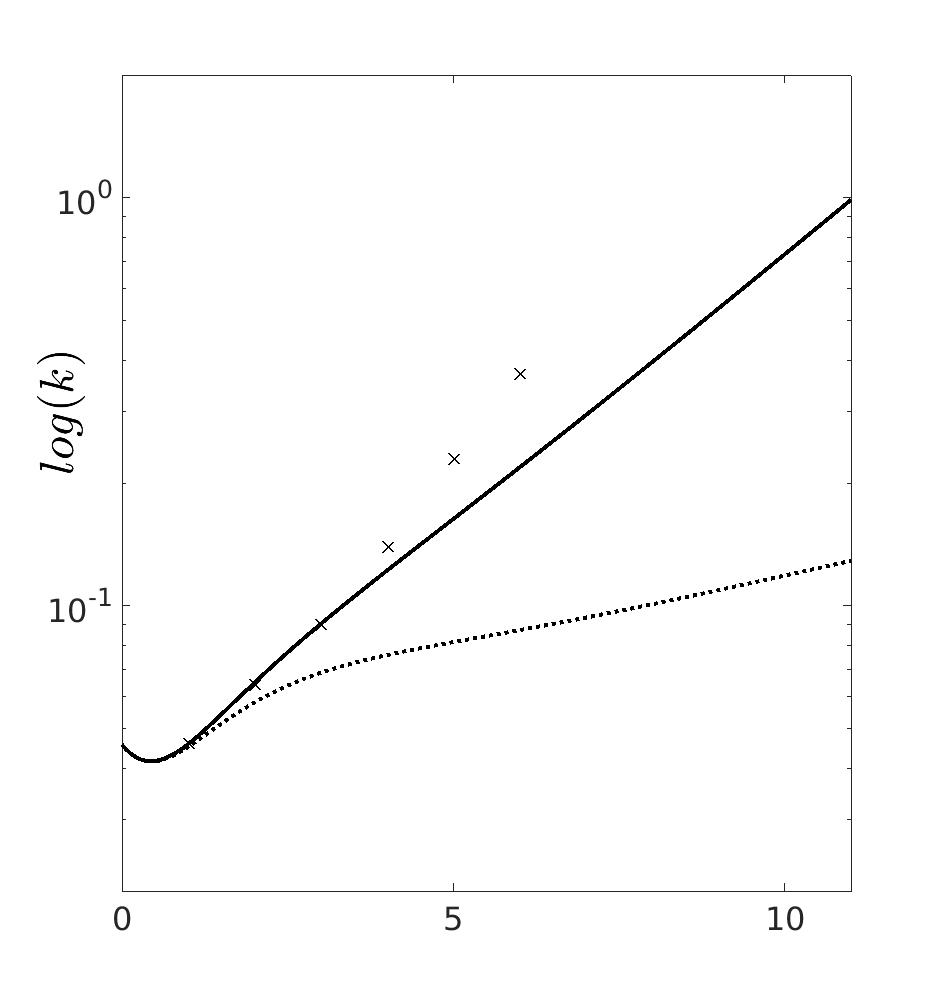}}
\caption{Rapid pressures train correlation model predictions for an elliptic streamline mean flow at the rapid distortion limit}
\end{figure}

\begin{figure}
\centering
\subfloat[$b_{22}$]{\includegraphics[height=8cm]{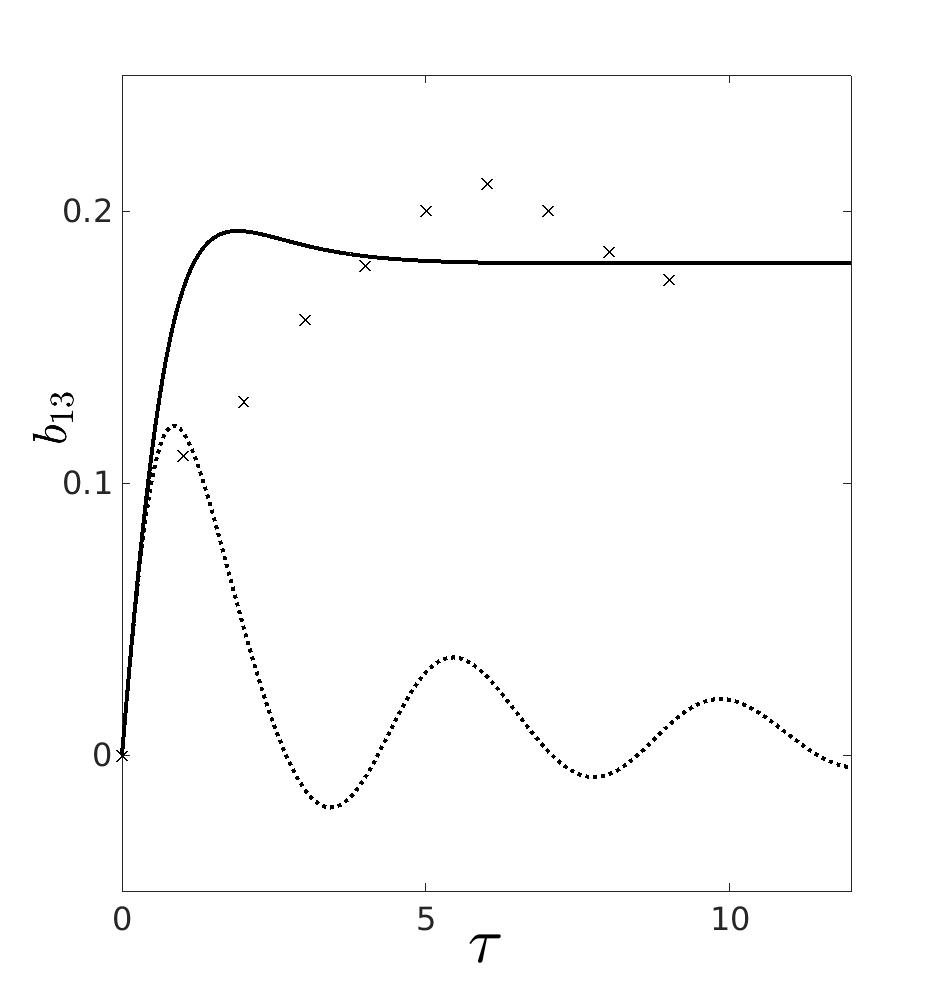}}
\subfloat[$log(k)$]{\includegraphics[height=8cm]{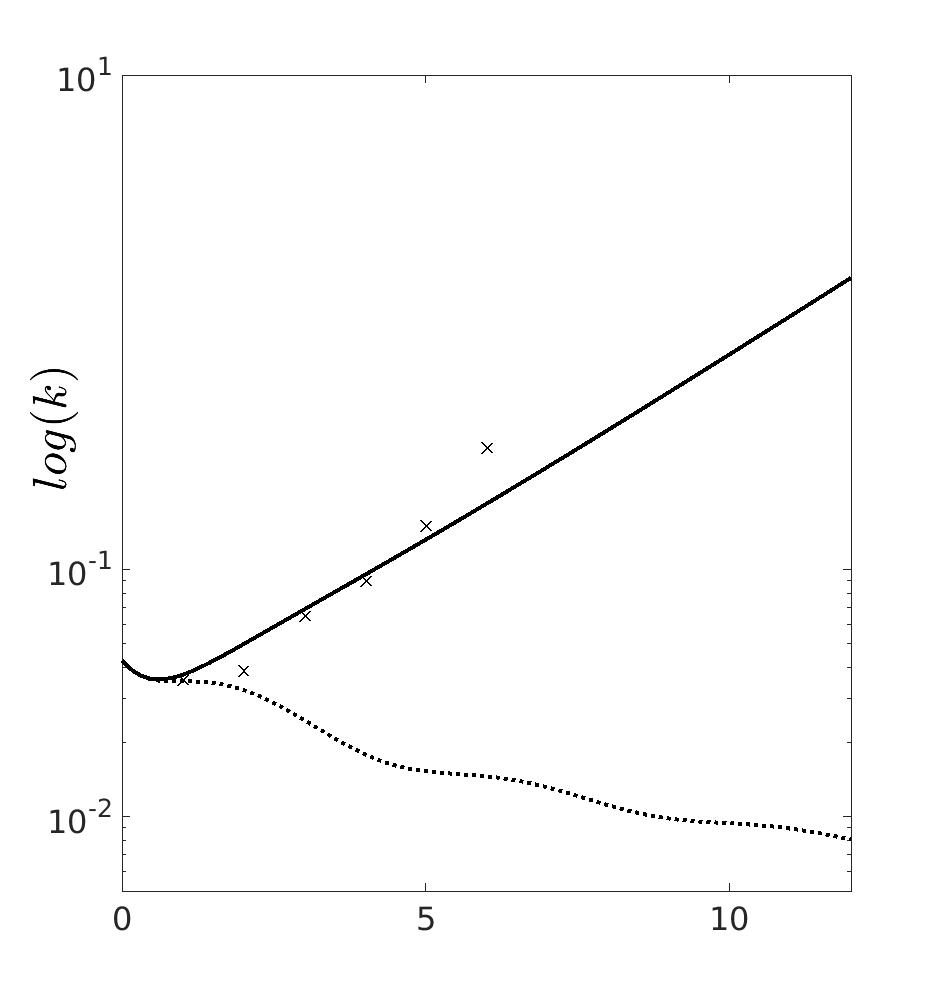}}
\caption{Rapid pressures train correlation model predictions for an elliptic streamline mean flow at the rapid distortion limit}
\end{figure}

\section{Modeling challenges and directions}
In this section of the review we outline and discuss some of the important present challenges and future research directions in pressure strain correlation modeling. It is our intention to make researchers cognizant of the details of these issues hampering the development and application of pressure strain correlation models in particular and the Reynolds stress modeling approach in general. 

\subsection{Quantification of errors, uncertainties and variability in predictions}
Owing to the simplifications made during the formulation of turbulence models they suffer from inaccuracies in their predictions. This uncertainty in the predictions can be epistemic or aleatoric. Epistemic uncertainties in turbulence modeling occur owing to the empirical nature of the closure modeling expressions, the inaccuracies in the closure modeling coefficients, etc \cite{uq3}. In the recent past uncertainty quantification for turbulence models has become a very active field where investigators are trying to estimate the discrepancy between model predictions and the true evolution of turbulence \cite{uq2, uq2}. Investigators have studied the nature of and the variability in the closure coefficients \cite{edeling}. The uncertainty arising due to the limitation of the closure expression have been investigated in detail by other researchers \cite{dow, smith}.

A vast majority of the study of uncertainty in turbulence models has been restricted to simpler 2-equation models like the $k-\epsilon$ and $k-\omega$ models. There has been very limited work done to quantify the uncertainties in and the variability of Reynolds stress modeling approaches. Reynolds stress models are as susceptible to such uncertainties and errors as zero-, 1- and 2-equation turbulence models. \cite{mishrauq2,mishrauq1} have studied the uncertainty in Reynolds stress model predictions arising due to the use of the Reynolds stress tensor as the only descriptor of the turbulent flow field. Using the Reynolds stress tensor as the only characteristic of the fluctuating velocity field ignores the history of the flow and assumes that all flows with the same Reynolds stresses will evolve identically under similar conditions. \cite{mishrauq2} have shown that this is not a satisfactory assumption and flows with the same Reynolds stress tensor can show evolution that is completely different. In light of these results it is essential to carry out careful studies analyzing the uncertainties in pressure strain correlation models. 

\subsection{Turbulent flows with compressibility effects}
In this article we have focused on the pressure strain correlation modeling for incompressible turbulent flows. Many flows of engineering interest have significant compressibility effects. These include high speed turbulent flows found in super-sonic and hyper-sonic design problems, turbulent flows in diffusers, etc. These flows are profoundly different from their incompressible counterparts \cite{cambon1993}. Compressible turbulent flows
have dilatational components of the velocity field leading to shocks, density variations and similar effects absent in incompressible flows \cite{sc}. Variability in the values of the transport coefficients that are functions of temperature and pressure are significant for compressible turbulent flows \cite{a1999}. The turbulent velocity field in compressible flows is highly dependent on the Mach number where the anisotropy of the turbulent flow increases with higher Mach numbers \cite{pantano}

It is accepted in the modeling community that majority of the compressibility effects are applied through the pressure strain correlation \cite{sarkar1}. The key hurdle in extending Reynolds stress modeling approach to compressible flows is the absence of an accurate model for the pressure strain correlation for these flows \cite{sarkar2}. While many investigators have studied pressure strain correlation modeling for incompressible flows there is much lesser work done for compressible flows. A majority of the pressure strain correlation models developed for compressible turbulent flows are extensions of available incompressible pressure strain correlation models with the addition of a blending function dependent on the turbulent Mach number \cite{c1,c2,c3,c4,c5,c6}. In numerical investigations it has been found that these models may give satisfactory results in very weakly compressible flows but in flows with significant compressibility their predictions are unsatisfactory \cite{gomez}. 

This approach of making small modifications to incompressible turbulent flow models for the pressure strain correlation has both advantages and disadvantages. As an advantage it is able to transition from weakly compressible to incompressible flows consistently. However it tends to ignore the significant changes in the nature of pressure between incompressible and compressible flows. In incompressible turbulent flows pressure acts as a Lagrange multiplier and ensures that the flow velocity remains divergence free. However in compressible flows pressure is a thermodynamic variable  governed  by the energy equation, the equation of state and calorific equation. At high levels of compressibility pressure has wave like behavior and this causes complex interactions with the
velocity field. This results in a significant change in the nature and evolution of the turbulent flow in compressible flows that is substantially different form the incompressible flows. The development of an accurate model for the pressure strain correlation that can capture the effects of compressibility in high speed flows and still exhibit consistency with incompressible physics in the limit of low speed flows is a key hurdle in turbulence modeling. 

\subsection{Improved modeling of the rate of dissipation tensor}
While models for the pressure strain correlation may be formulated in isolation their testing uses them with the rate of dissipation in simulating the flow. The coupled interaction of the pressure strain correlation model with the rate of dissipation model obfuscates the exact source of prediction errors. An important shortcoming for the Reynolds Stress Modeling approach is the approximate nature of the rate of dissipation equation. While the model equations for the evolution of the Reynolds stress anisotropy components are exact and based on the Reynolds stress transport equation the evolution equation for the rate of dissipation is empirically derived \cite{pope2000}. This model expression is 
\begin{equation}
\frac{D\epsilon}{Dt}=\frac{\partial }{\partial x_i} (\nu +\frac{\nu_t}{\sigma_k})\frac{\partial \epsilon}{\partial x_i} +C_{\epsilon 1} \frac{P\epsilon}{k} - C_{\epsilon 2} \frac{\epsilon^2}{k}
\end{equation}
The first term on the right hand side represents the diffusive transport of $\epsilon$. The second and third terms on the right side represent the generation of $\epsilon$ due to vortex stretching and the destruction of $\epsilon$ by viscous action. The standard values for the closure coefficients are given by $\sigma_{\epsilon}=1.3$, $C_{\epsilon 1}=1.44$ and $C_{\epsilon 1}=1.92$, based on the constants determined by \cite{launder1974}. The value of the $C_{\epsilon 2}$ coefficient is calibrated to be in agreement with the power law decay observed in decaying turbulence. Here the decay exponent corresponds to the power law decay observed as $k(t)=k(t_0)(t/t_0)^{-n}$ and $\epsilon(t)=\epsilon(t_0)(t/t_0)^{-n-1}$. In terms of the decay exponent $n$ this is given by
\begin{equation}
n=\frac{1}{C_{\epsilon 2} -1}
\end{equation}

\begin{figure}
\begin{centering}
\includegraphics[width=0.7\textwidth]{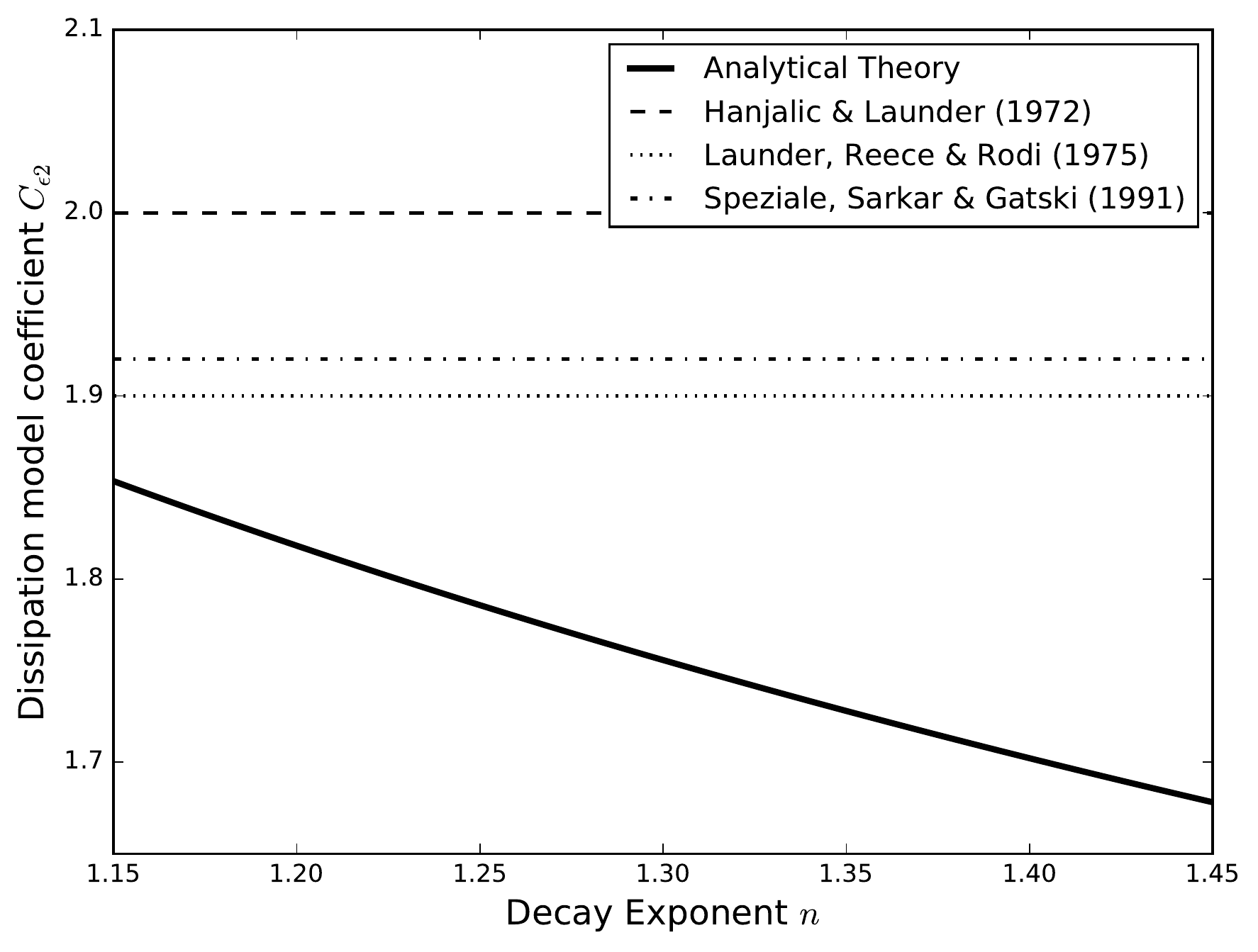}
\caption{Contrasting the relationship between $C_{\epsilon 2}$ and $n$ based on experimental studies (solid black line) and the values used in Reynolds Stress Modeling investigations. \label{fig:ce2}}
\end{centering}
\end{figure}

Most experimental investigations have found the decay exponent to lie in the range of $1.15-1.45$. This obligates the value of $C_{\epsilon 2}$ to approximately lie in the range $1.69-1.87$. However the values used in different models often lies well outside this bound. Based on \cite{batchelor1948}, \cite{hanjalic1972} chose $C_{\epsilon 2}=2.0$ to make the turbulent kinetic energy vary inversely with distance from the origin. Both the investigations of \cite{lrr} and \cite{ssmodel} changed it to $C_{\epsilon 2}=1.9$ so as to get faster rate of decay for their model simulations. \cite{ssg} chose to adopt the value of $C_{\epsilon 2}=1.92$ for better calibration of their model. Since then, different modeling investigations have used different values for the coefficient varying from $1.90$ to $2.0$. All these chosen values lie outside the range prescribed by experimental investigations and are often varying from one numerical investigation to another. 

The values of the $C_{\epsilon 1}$ is chosen to match the steady state parameters in homogeneous turbulent shear flow. The form is given by $\frac{P}{\epsilon}=\frac{C_{\epsilon 2} -1}{C_{\epsilon 1} -1}$. It can be seen from this relationship that the choice of the value of the coefficient $C_{\epsilon 2}$ also in turn affects the value of the $C_{\epsilon 1}$ coefficient. Any errors in the values of $C_{\epsilon 2}$ will have a cascading effect and will affect the accuracy of the entire model. As can be seen in this discussion there is significant inconsistency in the modeling of the rate of dissipation term. The errors due to the rate of dissipation term affect the evaluation of the pressure strain correlation models. Improved models for the rate of dissipation term may be of great assistance in developing and testing improved pressure strain correlation models.
\section{Concluding remarks}

In this article we provide a thorough review of pressure strain correlation modeling for turbulent flows. Starting from the Reynolds stress transport equations the numerical and mathematical foundations of pressure strain correlation modeling are established. The key challenges in this modeling effort arising due to the non-local nature of the behavior of the pressure strain correlation are established. 

Established slow pressure strain correlation models are introduced. Their predictions are compared and contrasted against experimental data from a range of experiments.

Popular rapid pressure strain correlation were introduced. Their predictions were contrasted against rapid distortion theory based simulations. It was shown that most rapid pressure strain correlation models have satisfactory behavior in strain dominated turbulent flows but unsatisfactory predictions in rotation dominated flows. Alternative models that add to the modeling basis were introduced and their predictions for elliptic streamline flows were shown. We outlined and discussed important challenges and hurdles for making pressure strain correlation models an indispensable tool in the engineering design process.

\newpage

\bibliographystyle{asmems4}
\bibliography{asme2e}

\end{document}